\documentclass[11pt,a4paper]{article}

\usepackage[a4paper,left=25mm,right=25mm,top=25mm,bottom=25mm]{geometry}
\usepackage{amsmath,amssymb,amsfonts}
\usepackage{graphicx}
\usepackage{booktabs}
\usepackage{xcolor}
\usepackage{float}
\usepackage{caption}
\usepackage{subcaption}
\usepackage{enumitem}
\usepackage{url}
\usepackage{cite}
\usepackage{array}
\usepackage{longtable}
\usepackage{multirow}
\usepackage{placeins}
\usepackage{microtype}
\usepackage{hyperref}

\hypersetup{
  colorlinks=true,
  linkcolor=blue,
  citecolor=blue,
  urlcolor=blue
}

\graphicspath{{figures/}{supp_figures/}}

\begin{document}
\sloppy

\begin{center}
{\LARGE\bfseries Electronic--morphological decoupling in mixed-dimensional\\ halide perovskite heterostructures}

\vspace{1.2em}

\begin{tabular}{c}
Mahmoud Samadpour\textsuperscript{a,b*},
Alireza Fathi Beiraghvandi\textsuperscript{a},
Kazem Meraji\textsuperscript{b},\\
Tristan Sachsenweger Ballantyne\textsuperscript{b},
Fuxiang Ji\textsuperscript{b},
Mahdi Mohammadi\textsuperscript{b},\\
Nasim Kabir\textsuperscript{b}
and Wolfgang Tress\textsuperscript{b*}
\end{tabular}

\vspace{0.8em}

{\small
\textsuperscript{a} Physics Department, K.~N.~Toosi University of Technology, Tehran, Iran\\
\textsuperscript{b} Institute of Computational Physics, Zurich University of Applied Sciences (ZHAW),\\
Winterthur 8400, Switzerland
}

\vspace{0.6em}

{\small *Corresponding authors email addresses: \texttt{wolfgang.tress@zhaw.ch}, \texttt{samadpour@kntu.ac.ir}}
\end{center}

\vspace{1em}

\begin{abstract}
Surface morphology is widely regarded as an important indicator of thin-film quality, with smoother films generally assumed to promote more uniform optoelectronic properties. Here, we demonstrate that marked structural evolution at the film surface does not necessarily lead to increased electronic heterogeneity in mixed-dimensional perovskite heterostructures. Specifically, we investigate whether the surface modifications introduced by low-dimensional (LD) Ruddlesden--Popper capping layers derived from the sterically hindered 5-tert-butyl-phenanthridinium cation are accompanied by corresponding changes in the spatial variations of photoluminescence (PL) characteristics and the extracted quasi-Fermi level splitting (QFLS) in 3D perovskites. By combining statistical analyses of atomic force microscopy with confocal, spatially resolved PL spectroscopy, we compare the evolution of topography with local optoelectronic properties at the microscale.

Depositing the capping layer increases surface roughness and height asymmetry; however, no corresponding increase in electronic inhomogeneity is observed. Additionally, the mixed-dimensional heterostructures exhibit enhanced photoluminescence, reduced non-radiative recombination, and significantly improved spatial uniformity of the radiative QFLS compared with the smoother pristine 3D films. Under accelerated aging conditions, the contrast between the increasingly disordered surface morphology and the preserved electronic consistency becomes even more pronounced. While pristine perovskite films develop substantial spatial fluctuations in emission characteristics and QFLS, the capped heterostructures largely preserve their homogeneity, proving that the LD layer effectively suppresses aging-induced optoelectronic disorder. This decoupling of electronic quality from surface irregularity provides a new design framework for engineering durable, high-performance perovskite photovoltaics.
\end{abstract}

\section{Introduction}

Perovskite solar cells (PSCs) have emerged as one of the most promising next-generation photovoltaic technologies owing to their rapid improvements in power conversion efficiency (PCE), now exceeding 27\%, and cost-effective fabrication methods [1]. However, despite these remarkable advancements, a major bottleneck preventing the commercialization of PSCs is their inherent instability. That is primarily attributed to moisture-induced decomposition, ion migration, and phase segregation in mixed-halide compositions [2]. Consequently, extensive research efforts have focused on improving the long-term stability of PSCs through compositional tuning, optimizing device architecture, passivation procedures, and interface engineering [3,4,5,6,7]. For instance, the use of mixed-cation compositions has been shown to suppress phase segregation and improve film crystallinity, leading to more stable PSCs [8,9,10]. Similarly, targeted halide or cation mixing (e.g., Br/I ratio tuning) can effectively enhance the resistance of perovskite films against ion migration and phase segregation [11]. Among the various interface engineering strategies, low-dimensional (LD) perovskite capping layers have emerged as an effective method for improving both stability and optoelectronic properties by suppressing defect-mediated recombination and protecting the underlying bulk perovskite layer.

LD perovskites have been widely explored for their superior moisture resistance, reducing ion migration, suppressing non-radiative recombination, and reducing surface defect density [12,13,14,15,16]. For instance, Grancini et al.\ highlighted how 2D perovskites serve as moisture barriers, extending the operational lifetime of PSCs [17]. A benzylamine-based spacer, namely 3,5-difluorobenzylamine (DF-BZA), was developed for stable and efficient quasi-2D Ruddlesden--Popper PSCs which exhibited superior film quality with significantly enlarged grain size and improved charge carrier lifetime compared to non-fluorinated BZA-based quasi-2D RP films [18]. Sun et al.\ systematically investigated the structural and photophysical properties of slot-die-coated Dion--Jacobson (DJ) and RP reduced-dimensional perovskites, revealing better photostability in DJ structures under operational conditions [19]. Further studies have revealed that LD layers also play a critical role in reducing ion migration, thereby suppressing hysteresis and enhancing the operational stability of PSCs under continuous illumination [19].

Studies utilizing steady-state PL and time-resolved PL (TRPL) measurements have indicated that LD layers lead to reduced recombination losses and enhanced carrier lifetime [20]. For example, Ghoreishi et al.\ observed that 2D/3D perovskite heterostructures exhibited enhanced PL intensities and lower defect densities compared to pristine 3D perovskites, suggesting the effective passivation of surface defects by LD layers [21].

Despite these advances, the formation of LD capping layers is frequently accompanied by substantial changes in surface morphology, including increased roughness, heterogeneous surface textures, and the emergence of secondary crystallites. Such morphological modifications are often viewed as a consequence of introducing bulky organic spacer cations and are commonly assessed when evaluating the quality of mixed-dimensional perovskite films. However, whether these structural changes are accompanied by corresponding variations in local optoelectronic properties remains largely unresolved.

In particular, it remains unclear whether the increased topographical complexity introduced by LD layers necessarily leads to enhanced electronic heterogeneity, or whether the improved optoelectronic properties commonly associated with LD capping layers can preserve a spatially coherent electronic behavior despite noticeable morphological disorder.

Addressing this question requires characterization techniques capable of directly probing local optoelectronic properties rather than relying solely on macroscopic device performance. Recent advances in spatially resolved microscopy and spectroscopy have highlighted the complex interplay of compositional variations, lattice strain, defect distributions, and interfacial chemistry in determining the optoelectronic properties of halide perovskites [22,23,24,25,26,27,28].

For example, cathodoluminescence imaging has resolved lateral bandgap variations in CsPbIBr$_2$ film [22], while wide-field hyperspectral imaging has mapped PSC electroluminescence and PL to estimate local QFLS through the generalized Planck's law [23]. Micro X-ray diffraction combined with time-resolved confocal PL imaging has revealed the impact of microscale strain variations on radiative recombination in perovskite films [24]. Recently, a multimodal operando microscopy toolkit was introduced to demonstrate how the microscale distributions of composition, recombination, and charge transport dictate both device performance and stability [25].

By combining quantitative optical spectroscopic techniques and synchrotron nanoprobe measurements, it was shown that nanoscale chemical heterogeneity dominates the optoelectronic response of alloyed PSCs [26]. Furthermore, using multidimensional PL imaging, physical parameters such as QFLS and Urbach energy were extracted at the micrometer scale to assess the passivation mechanism of phenylethylammonium-based cations in inverted PSCs [27]. Kelvin probe force microscopy and micro-PL mapping were also applied to probe the homogeneity of perovskite films treated by the ortho-isomer of (phenylene)di(ethylammonium) iodide (PDEAI$_2$) for tailored defect passivation [28].

These studies collectively highlight the importance of spatially resolved characterization for uncovering local structure--property relationships that remain inaccessible through conventional bulk measurements. Nevertheless, it remains unclear whether the substantial morphological changes introduced by mixed-dimensional architectures are necessarily accompanied by corresponding changes in microscale optoelectronic uniformity, particularly under aging conditions where spatial disorder becomes increasingly important.

In this study, we employ mixed-dimensional perovskite heterostructures to investigate the relationship between surface morphology and local optoelectronic properties. We introduce 5-tert-butyl-phenanthridinium iodide as a sterically hindered organic spacer that forms an RP capping layer on triple-cation mixed-halide perovskite films.

By combining statistical analyses of AFM images and spatially resolved photoluminescence measurements, we evaluate how morphological modifications relate to microscale optoelectronic coherence, including photoluminescence intensity and radiative QFLS extracted using the Lasher--Stern--W\"urfel formalism [29,30,31].

Our results reveal that, although the LD capping layer introduces distinct topographical disorder, it simultaneously enhances electronic homogeneity. This behavior is consistent with the effective suppression of non-radiative recombination, minimizing aging-induced local variations in QFLS.

More broadly, this work establishes that stable optoelectronic properties can be preserved despite substantial topographical variations, offering a new design paradigm for durable, high-performance perovskite photovoltaic interfaces.

\section{Experimental Section}

\subsection{Materials and methods}

All chemicals were purchased commercially and used as received without further purification. Titanium (IV) isopropoxide (TTIP, 99.99\%), methylammonium bromide (MABr, 99.5\%), formamidinium iodide (FAI, $>$99\%), lead bromide (PbBr$_2$, 99.99\%), cesium iodide (CsI, 99.99\%), 4-tert-butylpyridine (4-TBP, 99\%), lithium bis(trifluoromethanesulfonyl)imide (Li-TFSI, 99.95\%) and Spiro-OMeTAD [2,2$^\prime$,7,7$^\prime$-tetrakis-(N,N-di-p-methoxyphenylamine)-9,9$^\prime$-spirobifluorene, 99\%] were obtained from Sigma-Aldrich. Lead iodide (PbI$_2$, 99.99\%) was sourced from TCI. All solvents including 2-propanol (IPA, 99.5\%), ethanol ($>$99.8\%), acetonitrile (99.99\%), chlorobenzene (CB, 99.8\%), N,N-dimethylformamide (DMF, 99.9\%), dimethyl sulfoxide (DMSO, $>$99.9\%) were obtained from Sigma-Aldrich.

\subsection{Preparation of FTO/TiO$_2$/Perovskite Structures}

Here, FTO/Compact TiO$_2$/Mesoporous TiO$_2$/Perovskite layer (3D \& 3D/LD) structures were prepared as follows.

\subsubsection*{Deposition of compact and mesoporous TiO$_2$ Layer}
Fluorine-doped tin oxide (FTO) glass substrates (7--10~$\Omega$~sq$^{-1}$) were cleaned thoroughly using a soft brush with a diluted sodium hypochlorite solution. The cleaning process was followed by sonication in deionized water, acetone, isopropanol, and ethanol for 10~minutes each. After cleaning, the substrates were heated at 500~$^\circ$C for 1~hour to remove any residual organic impurities. A TiO$_2$ precursor solution was prepared by adding 0.021~g of HCl in ethanol (1~g) dropwise to a 0.178~g solution of TTIP dissolved in ethanol (1~g), under continuous stirring for 15~minutes. The mixture was then filtered using a 0.2~$\mu$m PTFE syringe filter. A compact TiO$_2$ layer was deposited on the FTO substrates by dynamic spin-coating (200~$\mu$l, 2000~r.p.m., 30~s). The coated substrates were then sintered at 500~$^\circ$C for 30~minutes, followed by a 10-minute UV ozone treatment. A mesoporous TiO$_2$ layer was deposited by spin-coating a diluted TiO$_2$ paste (weight ratio of 2:11 in ethanol) at 4000~r.p.m.\ for 30~seconds. The film was sintered at 500~$^\circ$C for 30~minutes, forming a crystalline anatase phase with a particle size of 20--25~nm.

\subsubsection*{Deposition of Perovskite Layer}
Here a mixed-cation perovskite layer with the Cs$_{0.08}$FA$_{0.8}$MA$_{0.12}$\allowbreak Pb(I$_{0.88}$Br$_{0.12}$)$_3$ composition was deposited on the mesoporous TiO$_2$ layer. The precursor solution was prepared by adding 658~mg PbI$_2$, 215~mg FAI, 32~mg CsI, 68~mg PbBr$_2$, and 20.8~mg MABr to a solvent mixture of DMF and DMSO at a 4:1 ratio. The solution was stirred and heated at 50~$^\circ$C for 10~minutes to form a clear yellow mixture. The precursor (45~$\mu$l) was applied to the prepared mesoporous layer via a two-step spin-coating process: (i) 1000~r.p.m.\ for 10~seconds and (ii) 6000~r.p.m.\ for 25~seconds. During the second step, 250~$\mu$l of chlorobenzene was dripped onto the film as an antisolvent 5~seconds before the end of the spin process. The resulting films were annealed on a pre-heated hot plate at 100~$^\circ$C for 45~minutes, producing a dark brown perovskite layer.

To intentionally introduce and study the impact of topographical variation on electronic uniformity, LD layers were deposited using 5-tert-butyl-phenanthridinium iodide precursor solution (1~mg~ml$^{-1}$ and 5~mg~ml$^{-1}$ in IPA) atop the 3D perovskite layer via spin-coating (100~$\mu$l, 4000~r.p.m., 30~s), followed by annealing at 100~$^\circ$C for 10~minutes.

For LD-only films on FTO substrate, a precursor of 1~mmol/ml PbI$_2$ and 2~mmol/ml 5-tert-butyl-phenanthridinium iodide were prepared in a solvent mixture of DMF:DMSO (4:1). The solution was stirred and heated at 50~$^\circ$C for 10~minutes and then applied on the FTO substrate via a two-step spin-coating process: (i) 1000~r.p.m.\ for 10~seconds and (ii) 6000~r.p.m.\ for 25~seconds. During the second step, 250~$\mu$l of chlorobenzene was dripped onto the film as an antisolvent 5~seconds before the end of the spin process. The resulting film was annealed on a pre-heated hot plate at 100~$^\circ$C for 45~minutes for better crystallization.

\subsection{Methods for Characterization}

A multi-modal characterization strategy was employed to specifically decouple the physical surface structure from the underlying electronic properties. The morphology of the 3D and 3D/LD layers was examined using a focused ion beam instrument (TFS Helios 600i machine) in TLD-SE mode with a 1000~V stage bias (effective landing energy 300--500~V) to maximize surface sensitivity. To rigorously quantify the topographical disorder introduced by the LD layer, AFM images were acquired with a Horiba LabRam HR Evolution in contact mode, allowing for the statistical calculation of RMS roughness and skewness. Crystallographic properties were analyzed using X-ray diffraction (XRD) with Cu~K$\alpha$ radiation on a Bruker D8 Advance diffractometer. Steady-state PL spectra were measured by means of a high-resolution spectrometer (Kymera 193i, Oxford Instruments) and a 473~nm laser (Coherent OBIS LX 473~nm) as excitation source. The absorption spectrum of the films was taken with the same spectrometer and a stabilized tungsten-halogen light source (SLS201L, Thorlabs). Time-resolved photoluminescence was recorded with a custom-built system using the TCSPC method (TimeHarp 260 board and PMA Hybrid 50 detector, PicoQuant). A 473~nm laser (FIU-15 supercontinuum white-light fiber laser and CONTRAST-SR-VIS tunable filter, NKT) with 1~MHz repetition rate was used as excitation. Central to our analysis of electronic homogenization, micro-scale PL mapping was performed to extract the spatial distribution of the quasi-Fermi level splitting via the Lasher--Stern--W\"urfel (LSW) relation. The pixel spacing in the PL maps was 3~$\mu$m.

For stability analysis, a climate chamber (BINDER, KMF 240) was used to simulate environmental conditions, enabling us to track how the relationship between topography and electronic uniformity evolves during the aging process.

\section{Results and discussion}

\begin{figure}[htbp]
\centering
\includegraphics[width=0.55\textwidth]{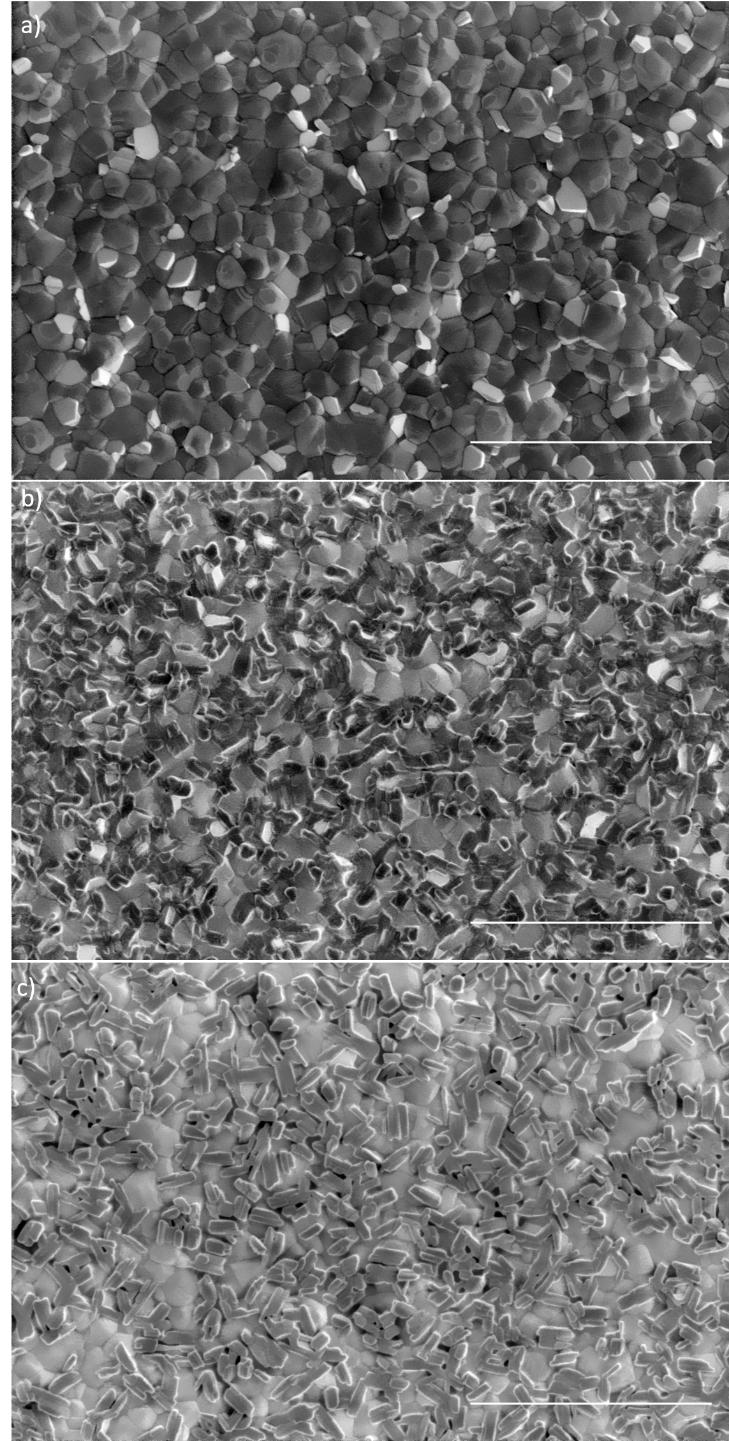}
\caption{SEM micrographs of (a) pristine 3D perovskite film, and 3D/LD structures treated with 5-tert-butyl-phenanthridinium iodide in IPA precursor solutions at concentrations of (b) 1~mg~ml$^{-1}$ and (c) 5~mg~ml$^{-1}$. The scale bar represents 2~$\mu$m for all pictures.}
\label{fig:SEM}
\end{figure}

Representative SEM micrographs of the pristine 3D and 3D/LD perovskite films are shown in Figure~\ref{fig:SEM}. The pristine triple-cation perovskite forms a compact, pinhole-free film composed of relatively large crystalline grains separated by well-defined grain boundaries. Following deposition of the 5-tert-butyl-phenanthridinium iodide solution, the surface morphology changes substantially. Distinct rod-like crystallites appear on the film surface, and their density increases with increasing precursor concentration from 1 to 5~mg~ml$^{-1}$, producing progressively more textured surfaces.

These observations indicate that formation of the RP phase introduces significant topographical intricacy without disrupting the continuity of the underlying perovskite film.

\begin{figure}[htbp]
\centering
\includegraphics[width=\textwidth]{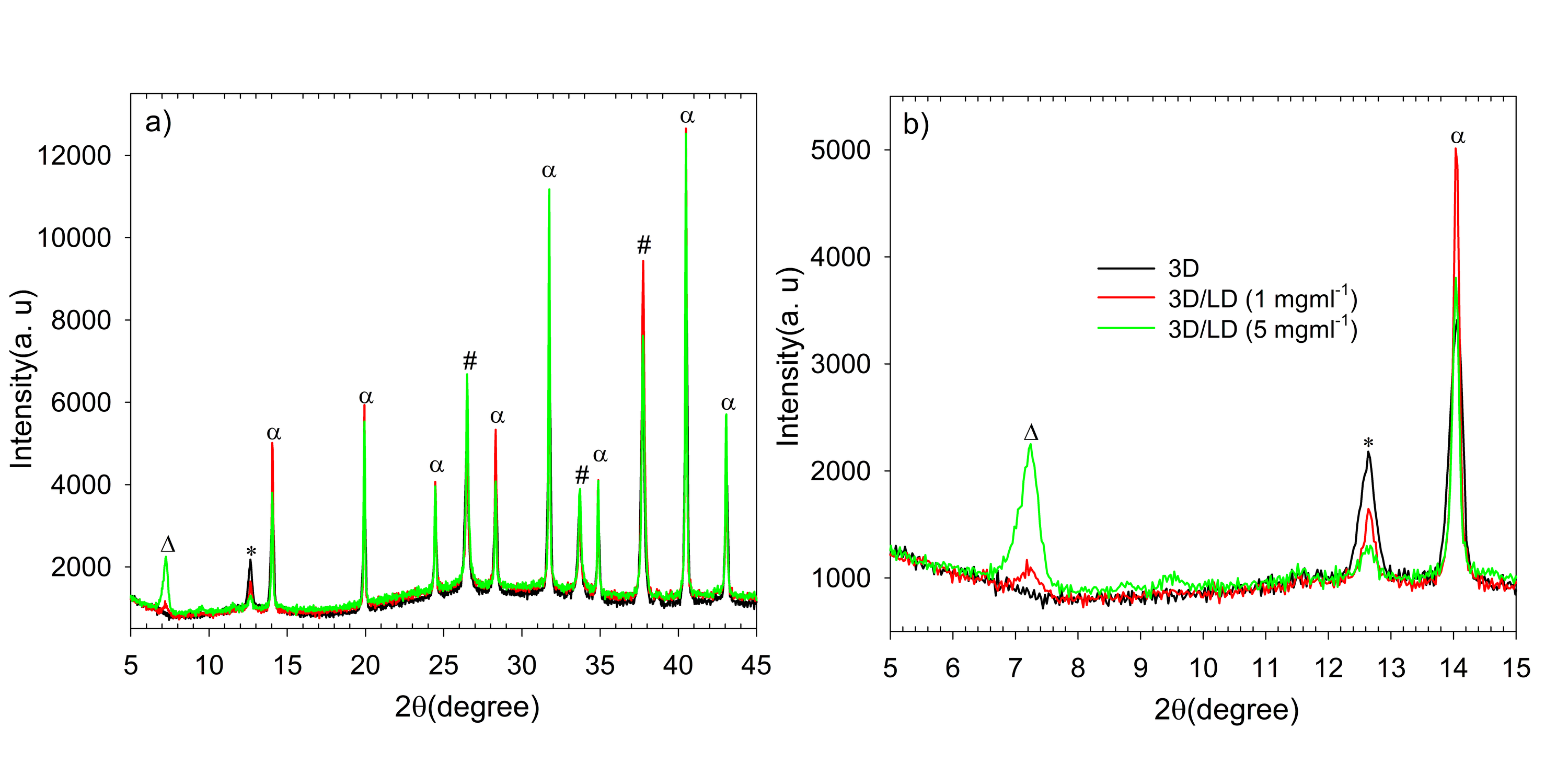}
\caption{(a) XRD patterns of pristine 3D and mixed-dimensional 3D/LD perovskite films deposited on FTO substrates. (b) Magnified view of the low-angle region (5$^\circ$--15$^\circ$) highlighting the emergence of low-dimensional perovskite phases. Diffraction peaks corresponding to the 3D perovskite matrix, LD phase, FTO substrate, and residual PbI$_2$ are indexed by $\alpha$, $\Lambda$, \#, and * respectively.}
\label{fig:XRD}
\end{figure}

Figure~\ref{fig:XRD} reveals the X-ray diffraction pattern of 3D and 3D/LD layers deposited on FTO substrates. The peaks at 7.25$^\circ$, 12.63$^\circ$, 14.06$^\circ$, 19.96$^\circ$, 24.45$^\circ$, 26.50$^\circ$, 28.32$^\circ$, 31.80$^\circ$, 33.70$^\circ$, 34.90$^\circ$, 37.76$^\circ$, 40.49$^\circ$, and 43.05$^\circ$, denoted by $\alpha$, are related to the crystalline planes of the 3D perovskite layer [8]. Following the LD layer deposition, a clear peak at 7.2$^\circ$ (annotated by $\Lambda$) is evident which increases by raising the LD layer precursor concentration (Figure~\ref{fig:XRD}b). Beside the perovskite layer and FTO substrate diffraction peaks (annotated by \#), the PbI$_2$ peak at 12.7$^\circ$ is observed (denoted by *), signifying the presence of non-reacted lead iodide in both 3D and 3D/LD structures [32]. However, the intensity of the PbI$_2$ peak declines through depositing the 5-tert-butyl-phenanthridinium iodide precursor solution atop the 3D layer and annealing (Figure~\ref{fig:XRD}b). These structural observations confirm that the post-treatment produces mixed-dimensional heterostructures while preserving the crystallographic integrity of the underlying 3D absorber.

\begin{figure}[htbp]
\centering
\includegraphics[width=\textwidth]{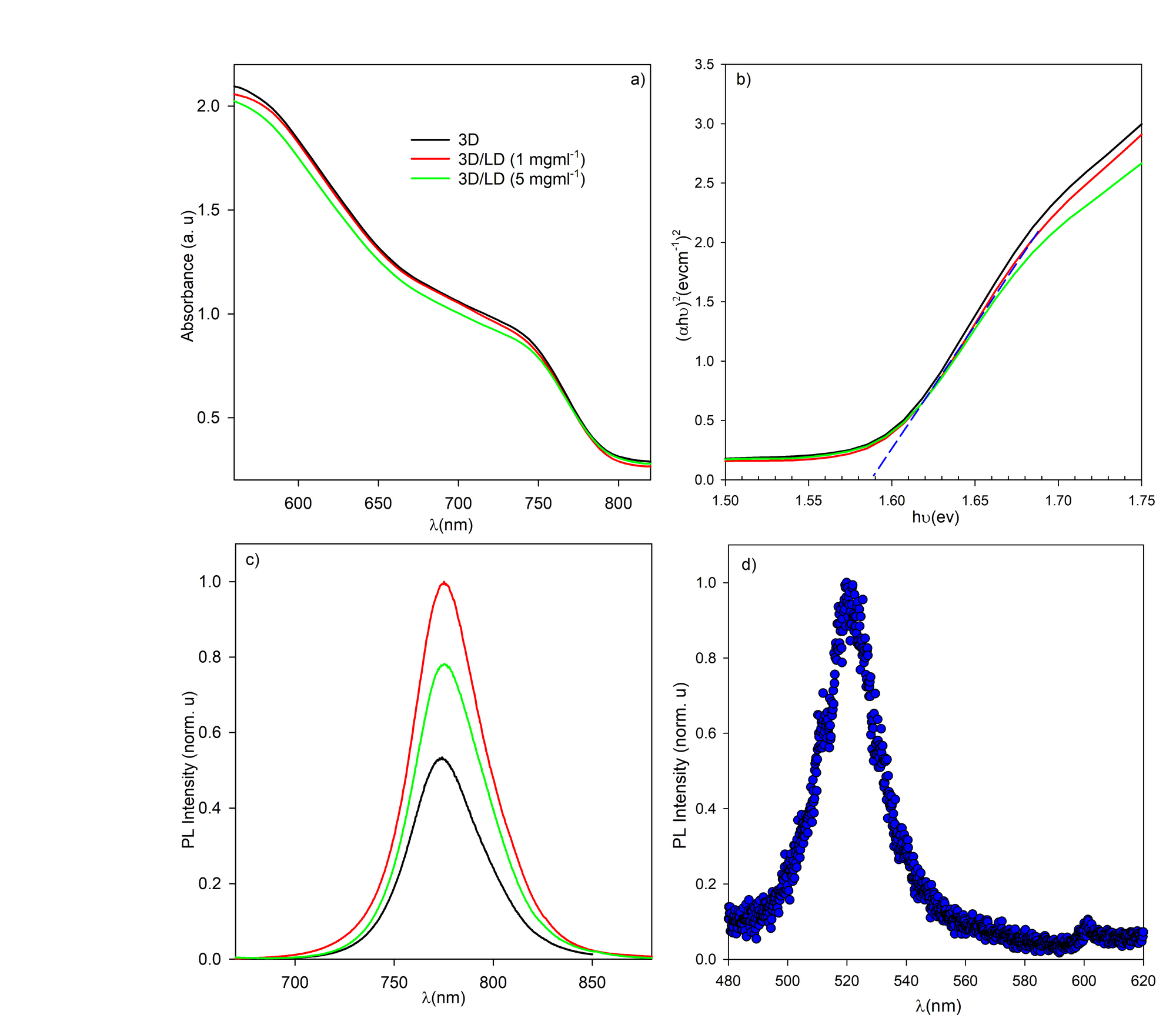}
\caption{Optical characterization of the perovskite films: (a) UV-Vis absorbance spectra, (b) corresponding Tauc plots for bandgap determination, (c) steady-state PL spectra of pristine 3D and 3D/LD films, and (d) steady-state PL spectrum of an isolated low-dimensional film deposited on an FTO substrate.}
\label{fig:optical}
\end{figure}

Figure~\ref{fig:optical}a and b present the optical absorption characteristics of the films. As shown in the UV-Visible absorption spectra (Figure~\ref{fig:optical}a), the absorption edge exhibits negligible variation after the LD layer deposition on the pristine 3D Cs$_{0.08}$FA$_{0.8}$MA$_{0.12}$\allowbreak Pb(I$_{0.88}$Br$_{0.12}$)$_3$ perovskite layer. Furthermore, the corresponding Tauc plot analysis (Figure~\ref{fig:optical}b) confirms that the optical bandgap values remain highly comparable at approximately 1.59~eV for both the pristine 3D and 3D/LD structures. This demonstrates that the electronic band structure of the bulk perovskite is largely preserved.

Likewise, the steady-state PL spectra (Figure~\ref{fig:optical}c) show almost identical emission wavelengths for all samples, suggesting that incorporation of the RP layer does not alter the emission energy of the underlying absorber. Instead, the primary effect of the RP treatment is a substantial enhancement of the PL intensity, consistent with suppression of non-radiative recombination through effective interfacial defect passivation.

The emission spectrum collected from the isolated RP film deposited on FTO (Figure~\ref{fig:optical}d) further confirms successful formation of the low-dimensional phase, dominated by the characteristic emission centered near 520~nm.

\begin{figure}[htbp]
\centering
\includegraphics[width=0.7\textwidth]{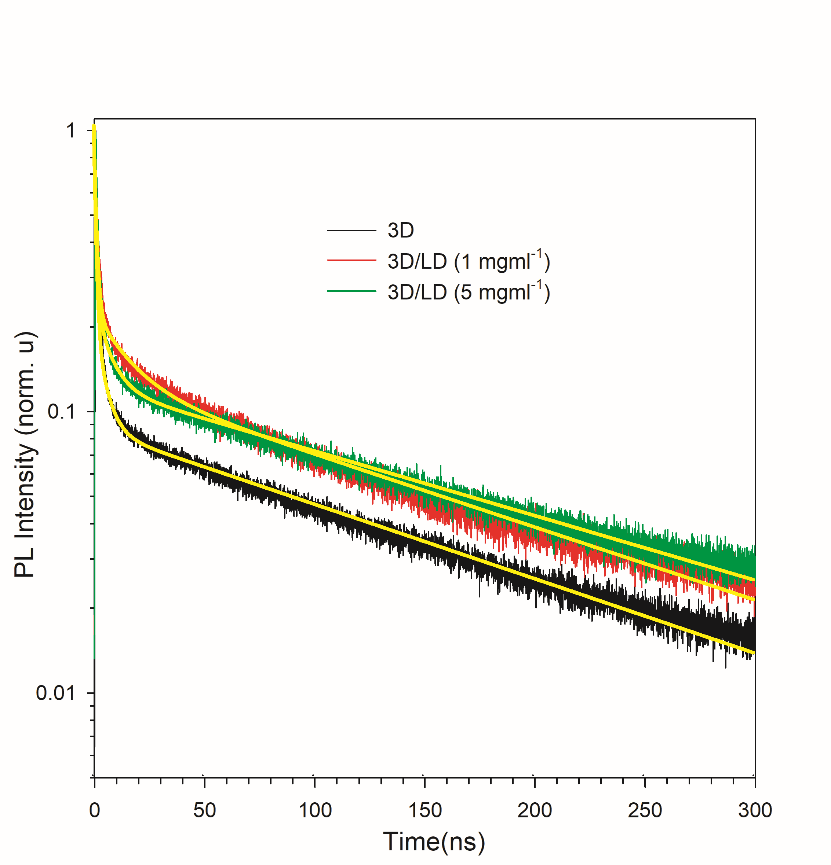}
\caption{Time-resolved photoluminescence decay profiles of pristine 3D and mixed-dimensional 3D/LD perovskite thin films. Solid yellow lines represent numerical fits obtained via a triexponential decay function.}
\label{fig:TRPL}
\end{figure}

The influence of the LD layers on charge carrier dynamics was further evaluated using TRPL (Figure~\ref{fig:TRPL}). Decay curves fitted using a triexponential function expressed as:
\begin{equation}
I(t) = A_1\exp(-t/\tau_1) + A_2\exp(-t/\tau_2) + A_3\exp(-t/\tau_3),
\end{equation}
where $I(t)$ is the PL intensity at time $t$; $A_1$, $A_2$, and $A_3$ are the amplitudes of the decay components; $\tau_1$, $\tau_2$, and $\tau_3$ are time constants to phenomenologically fit the experimental data. The average decay times were calculated by the amplitude-weighted average lifetime equation:
\begin{equation}
\tau_{\mathrm{avg}} = \frac{A_1\tau_1^2 + A_2\tau_2^2 + A_3\tau_3^2}
{A_1\tau_1 + A_2\tau_2 + A_3\tau_3}.
\end{equation}

The TRPL decay parameters are shown in Table~S1. Analysis of the decay curves yields average carrier lifetimes of 145, 149, and 174~ns for the pristine 3D film and the 3D/LD films prepared using 1 and 5~mg~ml$^{-1}$ precursor solutions, respectively.

The increase of $\tau_{\mathrm{avg}}$ from 145~ns to 174~ns suggests a slight suppression of non-radiative losses and thus improved charge carrier lifetime when the low-dimensional perovskite layer is introduced atop the 3D layer, consistent with the enhanced steady-state PL. These ensemble measurements therefore establish an important observation: although the LD capping layer produces intricate surface morphology, it simultaneously improves the average optoelectronic properties of the films. Whether these improvements remain spatially consistent across the film, however, cannot be determined from ensemble measurements alone and requires microscale characterization.

To investigate the spatial distribution of the optoelectronic response, PL mapping was performed on fresh and aged films after exposure to 70\% relative humidity and 35~$^\circ$C for a week in the dark.

\begin{figure}[htbp]
\centering
\includegraphics[width=0.95\textwidth]{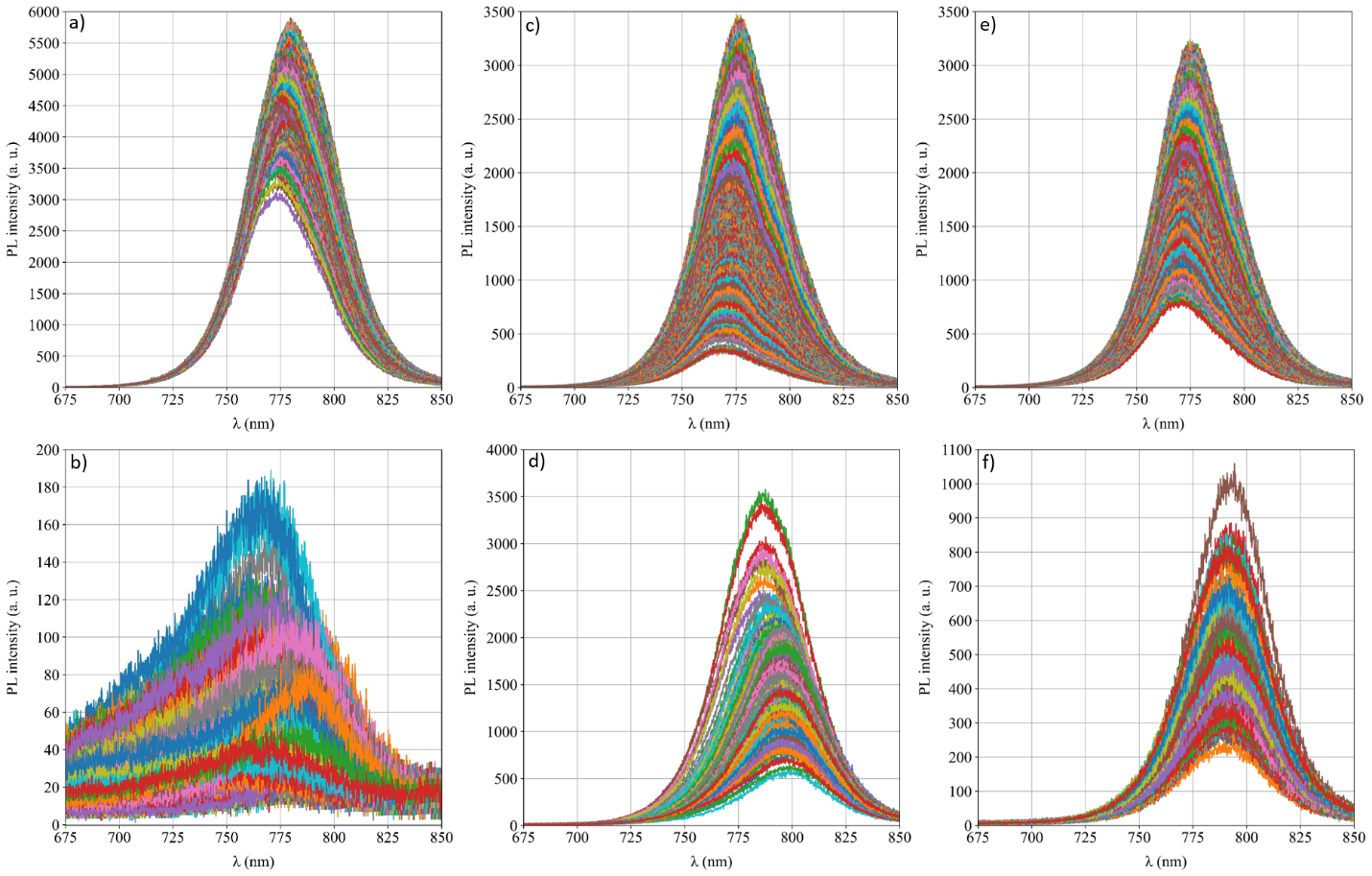}
\caption{Photoluminescence spectra extracted from the mapped regions: pristine 3D films in the (a) fresh and (b) aged states; 3D/LD (1~mg~ml$^{-1}$) films in the (c) fresh and (d) aged states; and 3D/LD (5~mg~ml$^{-1}$) films in the (e) fresh and (f) aged states. Environmental aging was performed at 35~$^\circ$C and 70\% relative humidity for one week in the dark.}
\label{fig:PLspectra}
\end{figure}

Measurements were acquired over areas of $200\times200$~$\mu$m (4096 spectra) or $500\times500$~$\mu$m (1024 spectra), enabling the evolution of the emission spectra to be examined with micrometer-scale resolution. Representative PL spectra obtained from the mapped regions (Figure~\ref{fig:PLspectra}) reveal that the intensity of PL peaks after aging is reduced remarkably in pristine films while the full width at half maximum (FWHM) is increased noticeably.

In contrast, the 3D/LD heterostructures retain much higher PL intensity with only minor spectral changes, demonstrating that the low-dimensional layer effectively preserves the radiative properties of the perovskite during aging.

\begin{figure}[htbp]
\centering
\includegraphics[width=0.55\textwidth]{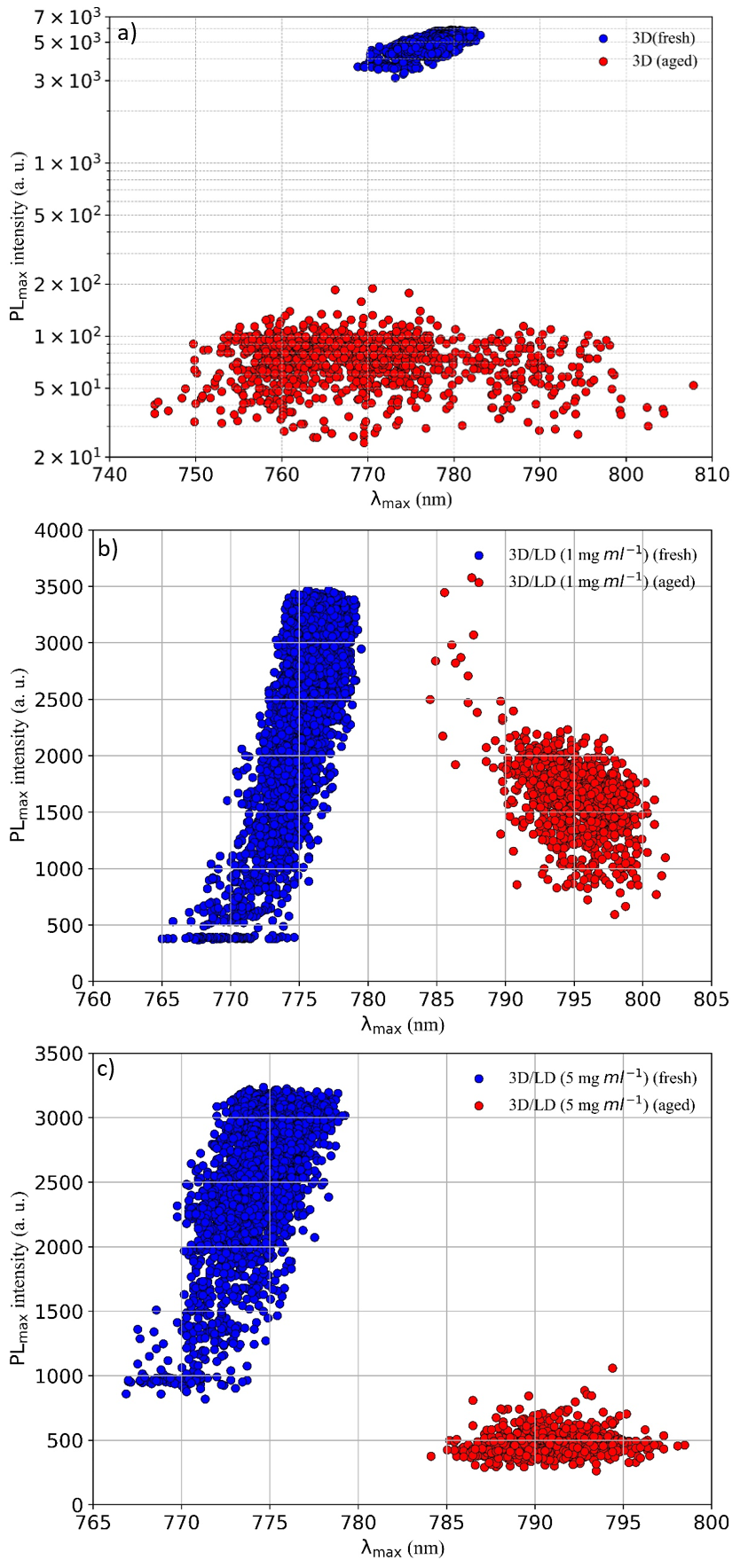}
\caption{Scatter plots of local peak PL intensity as a function of the emission maximum wavelength ($\lambda_{\max}$) for fresh (blue) and aged (red) thin films: (a) pristine 3D control matrix, (b) 3D/LD heterostructure (1~mg~ml$^{-1}$), and (c) 3D/LD heterostructure (5~mg~ml$^{-1}$).}
\label{fig:scatter}
\end{figure}

To get further insight, the scatter plots of PL intensities as a function of the local emission maximum ($\lambda_{\max}$) are presented in Figure~\ref{fig:scatter}, while the corresponding spatial maps of PL intensities and $\lambda_{\max}$ are plotted in Figures~S1 and S2 respectively. Fresh samples exhibit relatively narrow wavelength distributions irrespective of surface treatment, indicating initially homogeneous emission characteristics. After aging, however, the pristine perovskite develops a markedly broader $\lambda_{\max}$ distribution extending over approximately 750--810~nm, reflecting the emergence of spatially non-uniform degradation. This could be caused by moisture-induced halide segregation, leading to regions with lower-energy (iodide-rich) and higher-energy (bromide-rich) domains [33,34,35].

By contrast, the 3D/LD heterostructures preserve a narrow wavelength distribution even after aging, with their emission maxima remaining confined within a tight spectral range of 785 to 805~nm, indicating that the LD capping layer effectively suppresses localized optoelectronic disorder by mitigating defect formation and halide segregation. Thus, rather than merely delaying degradation, the LD layer maintains a remarkably regular electronic landscape during environmental exposure.

\begin{figure}[htbp]
\centering
\includegraphics[width=\textwidth]{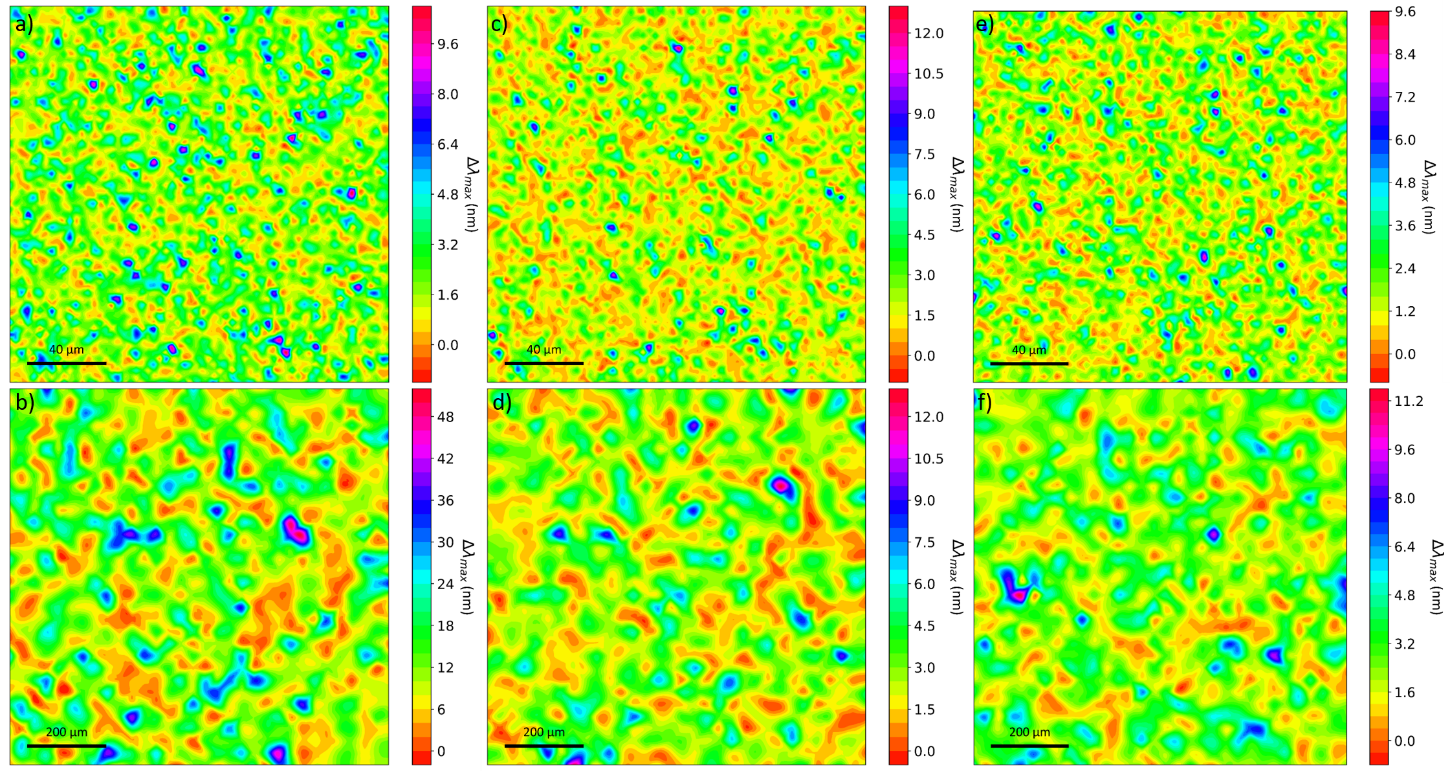}
\caption{Distribution maps of emission maximum variations ($\Delta\lambda_{\max}$) for: pristine 3D films in the (a) fresh and (b) aged states; 3D/LD (1~mg~ml$^{-1}$) films in the (c) fresh and (d) aged states; and 3D/LD (5~mg~ml$^{-1}$) films in the (e) fresh and (f) aged states. Interpolation was applied for noise reduction.}
\label{fig:Dlambda}
\end{figure}

To evaluate microscale optical uniformity, the average variation in peak wavelength ($\Delta\lambda_{\max}$) between each point and its four adjacent micron-spaced neighbors was analyzed (Figure~\ref{fig:Dlambda}).

In the fresh state, $\Delta\lambda_{\max}$ was below 12~nm for both 3D and 3D/LD layers. After aging, this value remained around 12~nm for 3D/LD layers, while it increased to approximately 50~nm in the 3D layers, highlighting the role of the LD capping layers in preserving optical homogeneity at the micro-scale.

For further analysis, the number of sites corresponding to each specific $\Delta\lambda_{\max}$ value was extracted, and the full width at half maximum (FWHM) of the resulting distribution was determined (Figure~S3). FWHM$_{\Delta\lambda_{\max}}$ values of 1.9, 1.7 and 1.8~nm were obtained for fresh 3D, 3D/LD (1~mg~ml$^{-1}$) and 3D/LD (5~mg~ml$^{-1}$) perovskite layers respectively. After aging these values increased to 16.8, 2 and 2.6~nm respectively. The large increase in FWHM$_{\Delta\lambda_{\max}}$ of 3D layers after aging, from 1.9 to 16.8~nm, suggests considerable degradation. On the other hand, modifying 3D layers with LD structures kept the FWHM$_{\Delta\lambda_{\max}}$ almost the same. The low FWHM$_{\Delta\lambda_{\max}}$ for 3D/LD layers after aging specifies superior stability, raised by 5-tert-butyl-phenanthridinium iodide precursor deposition atop the 3D layer.

While PL intensity offers valuable insights into radiative recombination efficiency, the QFLS provides a more direct, quantitative assessment of local electronic quality, prompting the use of the LSW equation [29,30,31] to analyze how aging impacts the spatial variation of the maximum obtainable QFLS in the radiative limit:
\begin{equation}
I_{\mathrm{PL}}(E) =
\frac{1}{4\pi^2 \hbar^3 c^2}
\, a(E) E^2
\left[
\exp\left(\frac{E-\Delta\mu}{k_{\mathrm{B}}T}\right)-1
\right]^{-1}
\end{equation}
Here $I_{\mathrm{PL}}$, $E$, $a(E)$, $\Delta\mu$, $k_{\mathrm{B}}$, and $T$ are PL intensity, photon energy, absorptivity, quasi-Fermi level splitting, Boltzmann constant, and carrier temperature respectively.
This equation describes the photoluminescence intensity as long as the absorption spectrum is exactly known. However, in the high-energy tail of the PL spectrum at photon energies sufficiently larger than the band gap, the spectral absorptivity $a(E)$ is approximately 1, simplifying the emission equation as follows:
\begin{equation}
\ln\left(\frac{4\pi^2 \hbar^3 c^2 I_{\mathrm{PL}}(E)}{E^2}\right)
= -\frac{E-\Delta\mu}{k_{\mathrm{B}}T}.
\end{equation}
To determine the temperature and QFLS, the quantity $\ln(I_{\mathrm{PL}}(E)/E^{2})$ was normalized and plotted against photon energy ($E$). In the high-energy region, this plot follows a linear trend, as the $I_{\mathrm{PL}}(E)/E^{2}$ exhibits an exponential dependence on energy. The slope of the fitted line corresponds to $-1/(k_{\mathrm{B}}T)$ allowing the extraction of the carrier temperature ($T$), while the extrapolated intercept provides information about separation of electron and hole quasi-Fermi levels ($\Delta\mu = E_{\mathrm{Fn}}-E_{\mathrm{Fp}}$). This approach allows for studying the aging effect on the perovskite layer's electronic properties on microscopic length scales, essential for evaluating its photovoltaic potential for long-term application.

\begin{figure}[htbp]
\centering
\includegraphics[width=0.75\textwidth]{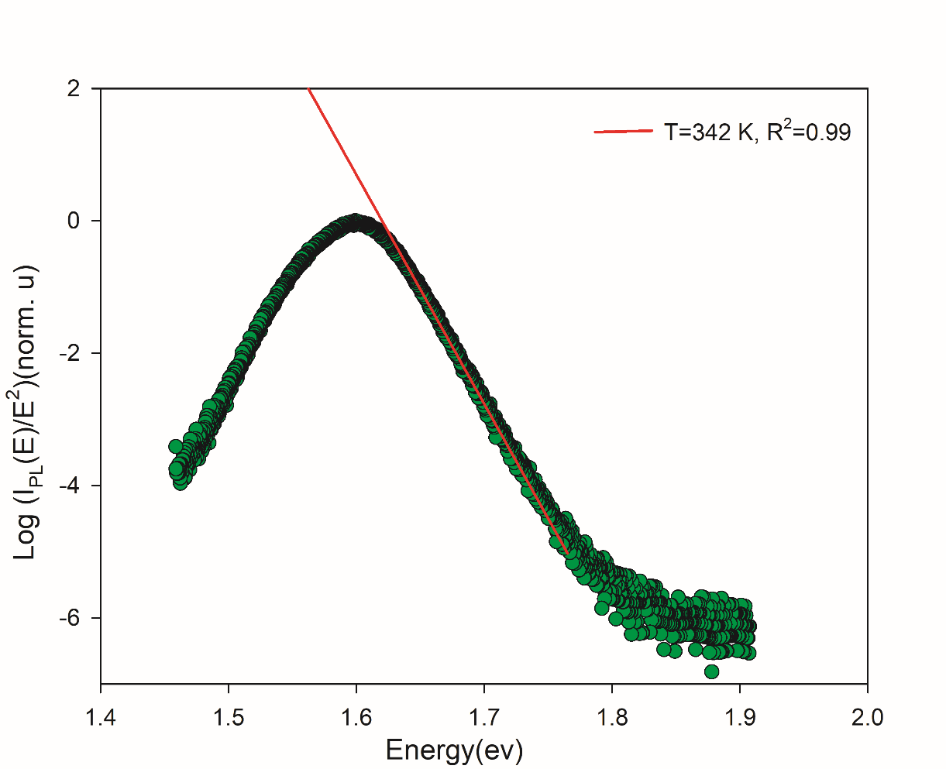}
\caption{Representative high-energy tail linear fit of $\ln(I_{\mathrm{PL}}(E)/E^{2})$ versus photon energy evaluated at a single spatial pixel within the PL hyperspectral map. The solid red line designates the linear regression fit ($R^{2}=0.99$) and extracted temperature of 342~K.}
\label{fig:LSWfit}
\end{figure}

Figure~\ref{fig:LSWfit} shows a representative plot of $\ln(I_{\mathrm{PL}}(E)/E^{2})$ vs photon energy, along with a fitted line in the high-energy region ($R^{2}=0.99$), yielding an extracted temperature of 342~K for a single pixel in the PL map. To analyze the spatial variation of the maximum obtainable quasi-Fermi-level splitting, such spectra are plotted for all 4096 pixels of any PL maps and fitted using the approach outlined earlier.

\begin{figure}[htbp]
\centering
\includegraphics[width=\textwidth]{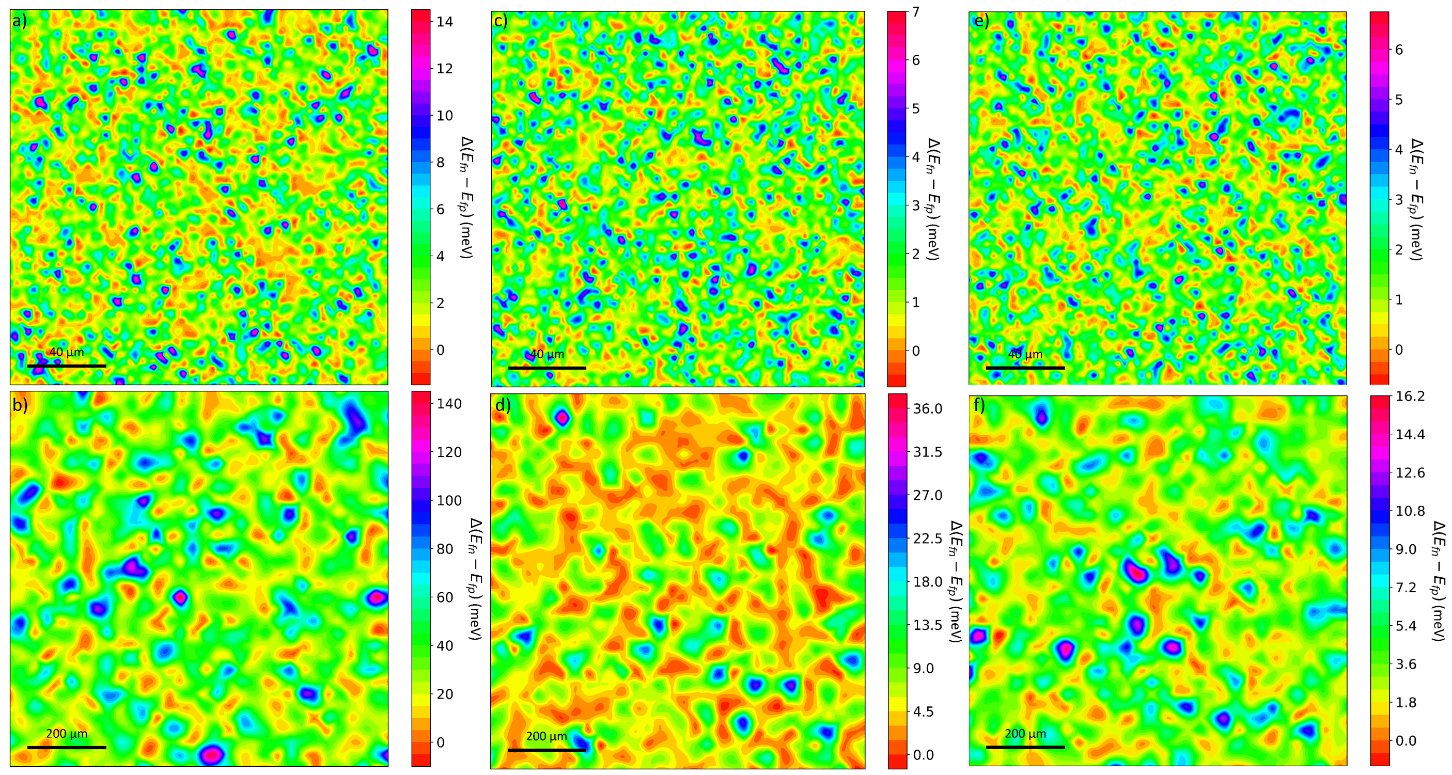}
\caption{Distribution maps capturing local radiative quasi-Fermi level splitting variations ($\Delta(E_{\mathrm{Fn}}-E_{\mathrm{Fp}})$) for: pristine 3D films in the (a) fresh and (b) aged states; 3D/LD (1~mg~ml$^{-1}$) films in the (c) fresh and (d) aged states; and 3D/LD (5~mg~ml$^{-1}$) films in the (e) fresh and (f) aged states. Accelerated degradation conditions correspond to 35~$^\circ$C and 70\% relative humidity in the dark. Interpolation was applied for noise reduction.}
\label{fig:DQFLS}
\end{figure}

As this experiment did not involve absolutely calibrated photon counts, we restricted ourselves to only analyzing the radiative contribution in the QFLS and neglecting contributions by varied PL yields [36]. Figure~\ref{fig:DQFLS} shows the average variation in QFLS ($\Delta(E_{\mathrm{Fn}}-E_{\mathrm{Fp}})$) between each point and its four adjacent micron-spaced neighbors for various samples, both before and after aging.

A striking observation in Figure~\ref{fig:DQFLS} is the decrease in the maximum of QFLS variation (from 14~meV in 3D to 7~meV in 3D/LD layers) after LD layer deposition, indicating that the low-dimensional perovskite layer improves the spatial homogeneity of the radiative recombination regime. This suggests better defect passivation and improved electronic coherence due to the presence of the LD layer. For a more precise analysis, the FWHM of QFLS variations (FWHM$_{\Delta(E_{\mathrm{Fn}}-E_{\mathrm{Fp}})}$) was derived from micro-PL measurements of 3D and 3D/LD perovskite films (Figure~S4).

Here the FWHM$_{\Delta(E_{\mathrm{Fn}}-E_{\mathrm{Fp}})}$ values were obtained as 2.7, 2.2 and 1.7~meV for fresh 3D, 3D/LD (1~mg~ml$^{-1}$) and 3D/LD (5~mg~ml$^{-1}$) perovskite layers respectively. After aging these values increased to 39.7, 6.6 and 4.1~meV respectively. The significant increase in the FWHM$_{\Delta(E_{\mathrm{Fn}}-E_{\mathrm{Fp}})}$ of the 3D layers after aging, from 2.7 to 39.7~meV, indicates marked chemical heterogeneity, likely resulting from moisture-induced halide segregation.

In contrast, the LD layer significantly suppresses this broadening (reaching only 4.1~meV for 5~mg~ml$^{-1}$), confirming its role in maintaining spatial compositional homogeneity.

Regarding the clear morphological modifications after LD layer deposition, the local fluctuations in the QFLS could originate from topographic features. So a statistical analysis was performed on AFM images of 3D and 3D/LD perovskite layers (Figure~S5) by Gwyddion software. The maximum height ($S_{\mathrm{z}}$) of the 3D, 3D/LD (1~mg~ml$^{-1}$), and 3D/LD (5~mg~ml$^{-1}$) samples were measured as 253, 272, and 315~nm, respectively. This indicates that the introduction of the LD layer increases the maximum height compared to the 3D layer. At higher precursor concentrations, the rod-like structures become more pronounced resulting in a significantly higher maximum height. To further analyze the surface morphology, the root-mean-square (RMS) roughness and skewness ($S_{\mathrm{sk}}$) of layers were investigated. While RMS roughness ($S_{\mathrm{q}}$) is commonly used to measure surface roughness, $S_{\mathrm{sk}}$ is a dimensionless parameter that describes the asymmetry of the surface height distribution. The RMS roughness values were 29.8~nm, 32.3~nm, and 34.6~nm, while the $S_{\mathrm{sk}}$ values were 0.11, 0.19, and $-0.05$ for the 3D, 3D/LD (1~mg~ml$^{-1}$), and 3D/LD (5~mg~ml$^{-1}$) layers, respectively. The $S_{\mathrm{sk}}$ value of 0.11 for the 3D layer indicates a surface with slightly more peaks than valleys, characteristic of the uniform grain morphology of 3D perovskites. In contrast, the higher $S_{\mathrm{sk}}$ value of 0.19 for the 3D/LD (1~mg~ml$^{-1}$) layer suggests the presence of more pronounced surface peaks, arising from the formation of rod-like structures on the 3D surface. These rods are unevenly distributed, resulting in a surface dominated by protrusions. For the 3D/LD (5~mg~ml$^{-1}$) layer, the negative $S_{\mathrm{sk}}$ value ($-0.05$) suggests a transition to a surface with deeper valleys between the more elongated rod structures. This shift from positive to negative skewness indicates increased aspect ratio without necessarily implying higher surface coverage.

Intriguingly, this morphological intricacy contrasts with the films' optoelectronic properties. While AFM reveals progressively broader height distributions, microscale PL measurements show narrower wavelength distributions, reduced QFLS fluctuations, and significantly improved electronic homogeneity. This independence indicates that the RP capping layer successfully passivates interfacial defects to suppress non-radiative recombination during aging, rendering the electronic profile largely independent of surface roughness. Consequently, these findings establish electronic homogeneity as an important design criterion for durable, high-performance mixed-dimensional perovskite interfaces, taking precedence over morphological smoothness. Given the diffraction-limited spatial resolution of PL measurements, future studies could combine tip-enhanced PL with co-localized AFM imaging to directly investigate the relationship between nanoscale morphology and radiative recombination.

\section{Conclusion}

In this work, we investigated the interplay between surface morphology and local optoelectronic properties in mixed-dimensional perovskite heterostructures engineered with an in-situ RP capping layer based on 5-tert-butyl-phenanthridinium iodide. Although the RP layer substantially increased surface roughness and structural intricacy, these topographical changes did not compromise the electronic quality. Instead, the mixed-dimensional heterostructures demonstrated prolonged carrier lifetimes, suppressed non-radiative recombination, and improved spatial uniformity in radiative QFLS.

Spatially resolved photoluminescence spectroscopy revealed that this beneficial impact is particularly vital under accelerated aging conditions. While the pristine three-dimensional perovskite developed distinct spatial fluctuations in emission and QFLS due to localized degradation, the RP-capped heterostructures largely preserved a homogeneous QFLS. Crucially, complementary AFM and microscale photoluminescence data demonstrated a distinct independence between electronic quality and surface irregularity. Instead of causing an increase in local electronic disorder due to the rougher surface, the RP capping layer successfully passivates interfacial defects, stabilizing the optoelectronic homogeneity despite a highly irregular surface topography.

More broadly, these findings highlight that spatially resolved characterization is indispensable for directly probing the electronic behavior, as it captures localized optoelectronic features that structural metrics cannot predict.

Ultimately, the concept of electronic--morphological independence validated here provides a valuable design framework for mixed-dimensional photovoltaics. It demonstrates that preserving electronic homogeneity, rather than solely optimizing morphological smoothness, represents a key priority for achieving both exceptional durability and device performance.

\section*{Acknowledgements}

The authors gratefully acknowledge the support from the Swiss Leading House South Asia and Iran and the European Research Council (ERC). They also thank ScopeM for their valuable support and assistance in this work.

% ---------- Bibliography ----------

\clearpage
\appendix
\section*{Supporting Information}

\renewcommand{\thefigure}{S\arabic{figure}}
\renewcommand{\thetable}{S\arabic{table}}
\setcounter{figure}{0}
\setcounter{table}{0}

\subsection*{Table S1. Fitting parameters for the time-resolved PL decay from 3D and 3D/LD perovskite films.}

\begin{table}[htbp]
\centering
\begin{tabular}{lccccccc}
\toprule
Sample & $A_1$ & $A_2$ & $A_3$ & $\tau_1$ (ns) & $\tau_2$ (ns) & $\tau_3$ (ns) & $R^2$ score \\
\midrule
3D perovskite & 1.35 & 0.14 & 0.08 & 0.8 & 4.7 & 163.9 & 0.99 \\
3D/LD (1~mg~ml$^{-1}$) & 1.16 & 0.10 & 0.12 & 1.1 & 15.2 & 167.6 & 0.99 \\
3D/LD (5~mg~ml$^{-1}$) & 1.27 & 0.13 & 0.12 & 0.9 & 6.8 & 189.2 & 0.99 \\
\bottomrule
\end{tabular}
\end{table}

\begin{figure}[htbp]
\centering
\includegraphics[width=0.95\textwidth]{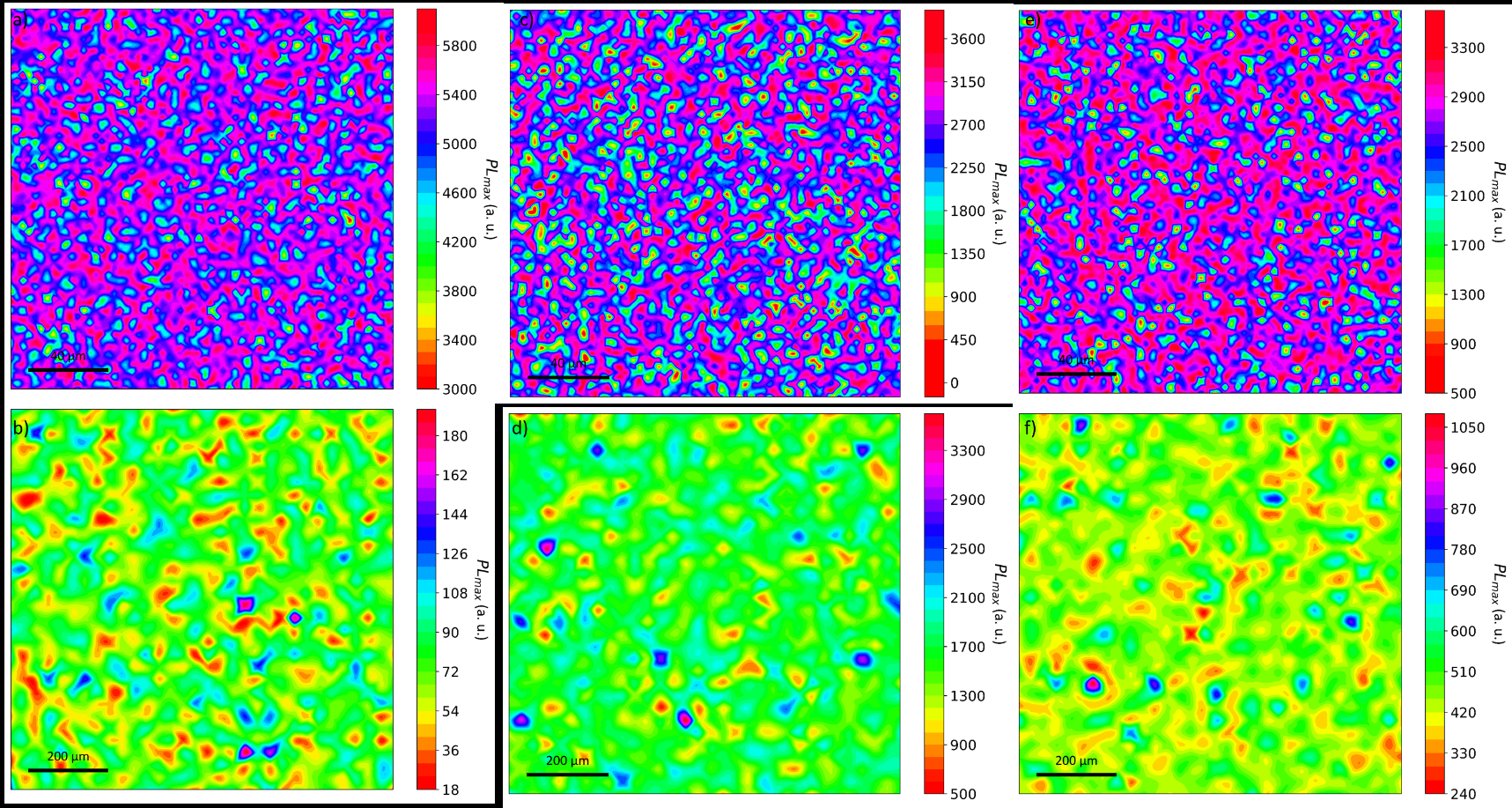}
\caption{Spatial PL intensity maps of pristine 3D perovskite films in the (a) fresh and (b) aged states; 3D/LD (1~mg~ml$^{-1}$) films in the (c) fresh and (d) aged states; and 3D/LD (5~mg~ml$^{-1}$) films in the (e) fresh and (f) aged states. Environmental aging was carried out under 70\% relative humidity at 35~$^\circ$C for one week in the dark.}
\label{fig:S1}
\end{figure}

\begin{figure}[htbp]
\centering
\includegraphics[width=0.95\textwidth]{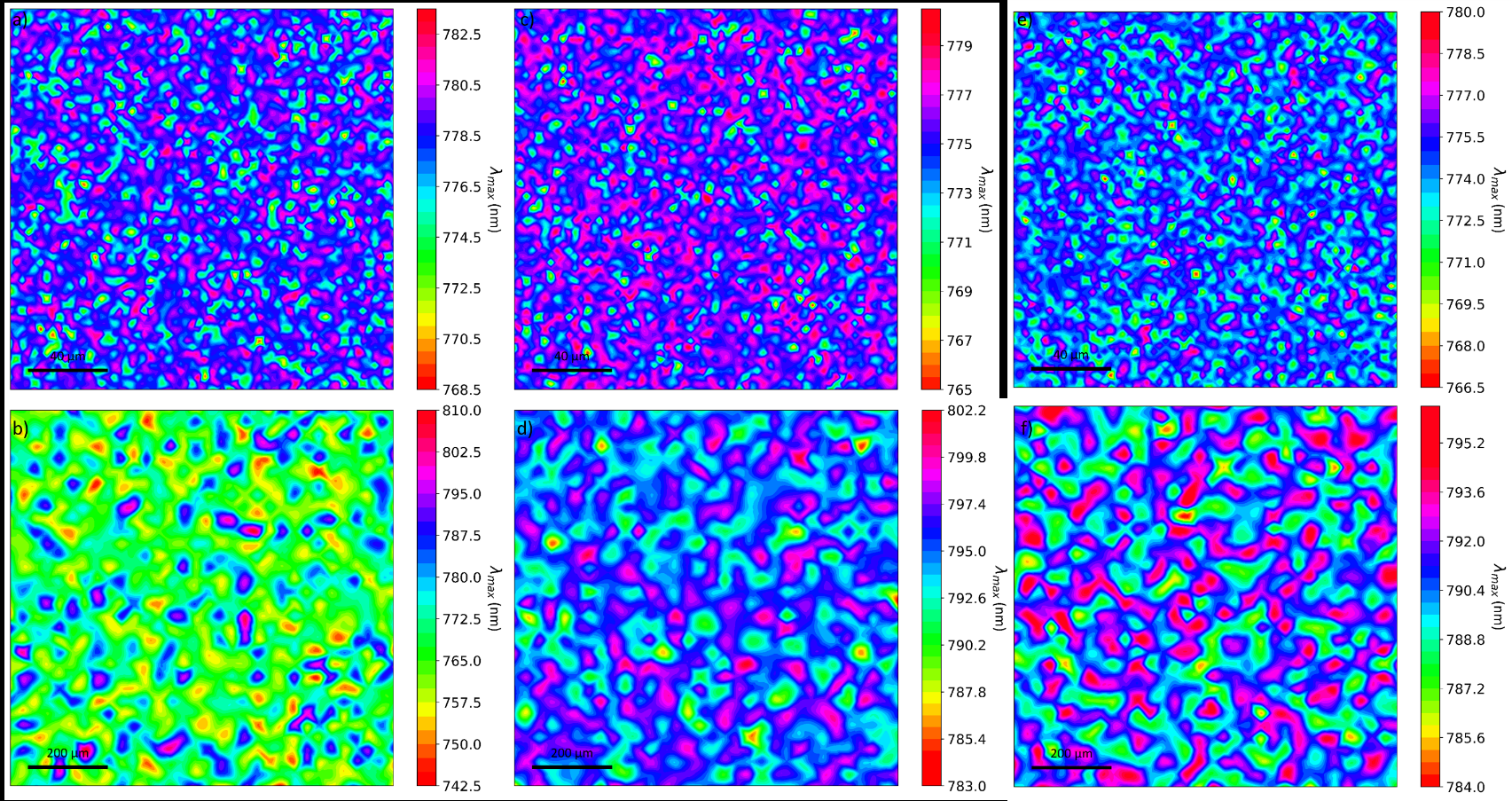}
\caption{Spatial distribution maps of the peak emission wavelength ($\lambda_{\max}$) for pristine 3D perovskite films in the (a) fresh and (b) aged states; 3D/LD (1~mg~ml$^{-1}$) films in the (c) fresh and (d) aged states; and 3D/LD (5~mg~ml$^{-1}$) films in the (e) fresh and (f) aged states. Environmental aging was carried out under 70\% relative humidity at 35~$^\circ$C for one week in the dark.}
\label{fig:S2}
\end{figure}

\begin{figure}[htbp]
\centering
\includegraphics[width=0.7\textwidth]{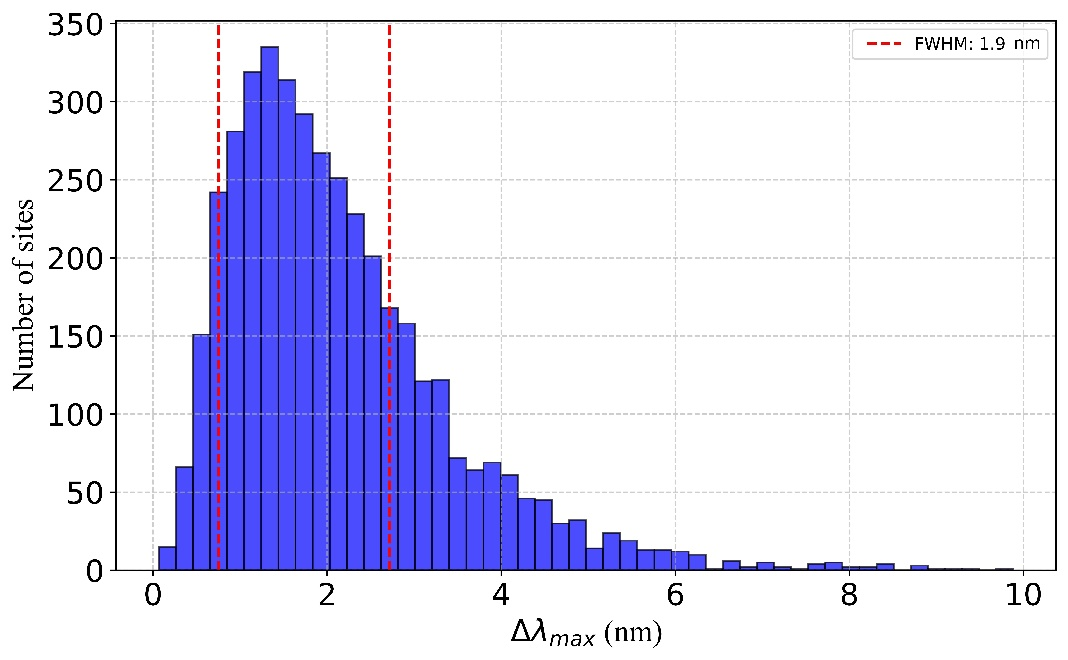}
\caption{Typical plot of number of sites at each specific $\Delta\lambda_{\max}$. The dashed red lines indicate the full width at half maximum (FWHM) of the corresponding abundance profile.}
\label{fig:S3}
\end{figure}

\begin{figure}[htbp]
\centering
\includegraphics[width=0.7\textwidth]{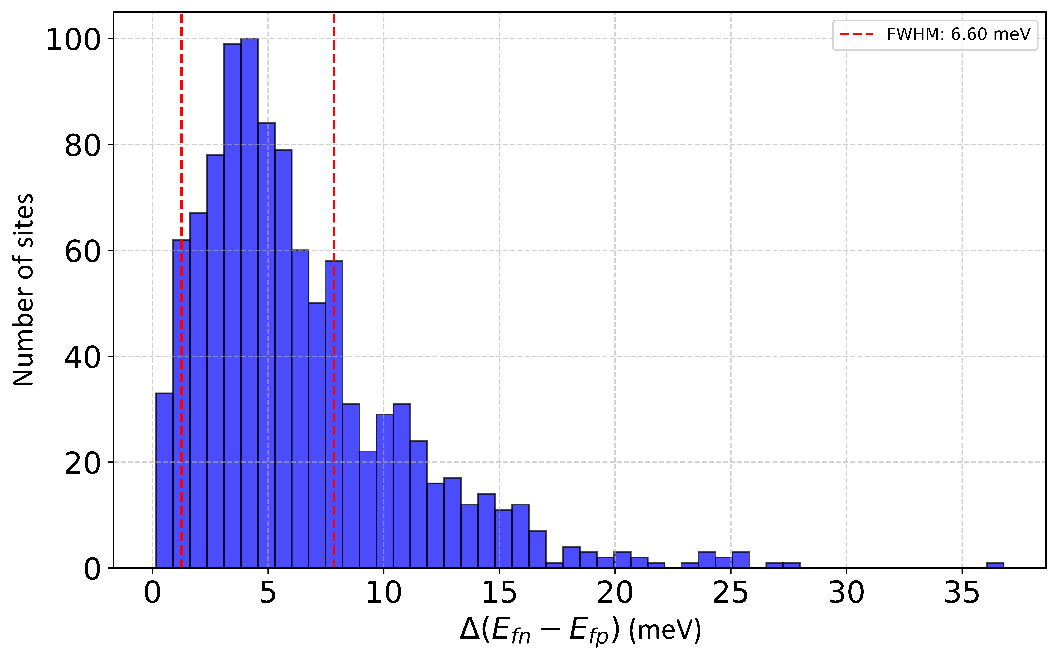}
\caption{Typical plot of number of sites at each specific $\Delta(E_{\mathrm{Fn}}-E_{\mathrm{Fp}})$. The dashed red lines indicate the FWHM of the corresponding abundance profiles.}
\label{fig:S4}
\end{figure}

\begin{figure}[htbp]
\centering
\includegraphics[width=0.55\textwidth]{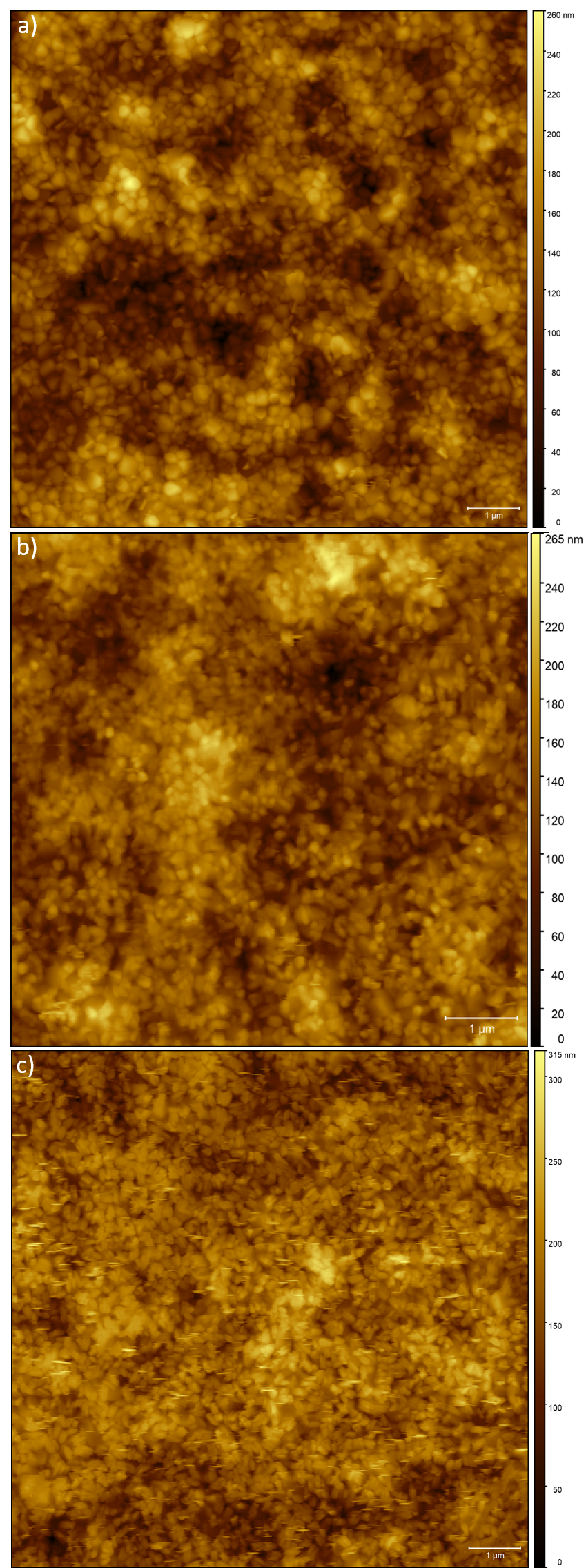}
\caption{AFM topography micrographs of (a) pristine 3D, (b) 3D/LD (1~mg~ml$^{-1}$), and (c) 3D/LD (5~mg~ml$^{-1}$) perovskite thin films. The scale bar represents 1~$\mu$m for all pictures.}
\label{fig:S5}
\end{figure}

\end{document}